\documentclass{article}
\usepackage{spconf,amsmath,graphicx}

\usepackage{subcaption}  % to replace subfig
\usepackage{booktabs}  % professional table
\usepackage{multicol}
\usepackage{array}

\usepackage{xcolor}

\usepackage[protrusion=true,expansion=true]{microtype}

\usepackage{amsmath,amsfonts,bm}
\usepackage{mathtools}
\usepackage{mathrsfs}

\def\eqref#1{equation~\ref{#1}}
\def\1{\bm{1}}

\DeclareMathAlphabet{\mathsfit}{\encodingdefault}{\sfdefault}{m}{sl}
\SetMathAlphabet{\mathsfit}{bold}{\encodingdefault}{\sfdefault}{bx}{n}

\def\gF{{\mathcal{F}}}

\def\gL{{\mathcal{L}}}

\def\gT{{\mathcal{T}}}

\def\sR{{\mathbb{R}}}

\newcommand{\R}{\mathbb{R}}

\def\t#1{\tilde{#1}}

\def\d#1{\dot{#1}}

\def\p{\partial}

\def\L#1{\left#1}
\def\R#1{\right#1}

\title{Speech-Based Parameter Estimation of an Asymmetric Vocal Fold Oscillation Model and Its Application in Discriminating Vocal Fold Pathologies}
\name{Wenbo Zhao$^{\dag}$ \quad Rita Singh$^{\ddag}$
\sthanks{
	Copyright 2020 IEEE. Published in the IEEE 2020 International Conference on Acoustics, Speech, and Signal Processing (ICASSP 2020), scheduled for 4-9 May, 2020, in Barcelona, Spain. Personal use of this material is permitted. However, permission to reprint/republish this material for advertising or promotional purposes or for creating new collective works for resale or redistribution to servers or lists, or to reuse any copyrighted component of this work in other works, must be obtained from the IEEE. Contact: Manager, Copyrights and Permissions / IEEE Service Center / 445 Hoes Lane / P.O. Box 1331 / Piscataway, NJ 08855-1331, USA. Telephone: + Intl. 908-562-3966.
}
}
\address{$^{\dag}$Department of Electrical and Computer Engineering, $^{\ddag}$School of Computer Science\\
Carnegie Mellon University, Pittsburgh, USA\\
\small \tt wzhao1@andrew.cmu.edu, rsingh@cs.cmu.edu}
\begin{document}
\ninept
\maketitle
\begin{abstract}
So far, several physical models have been proposed for the study of vocal fold oscillations during phonation. The parameters of these models, such as vocal fold elasticity, resistance, etc. are traditionally determined through the observation and measurement of the vocal fold vibrations in the larynx. Since such direct measurements tend to be the most accurate, the traditional practice has been to set the parameter values of these models based on measurements that are averaged across an ensemble of human subjects. However, the direct measurement process is hard to revise outside of clinical settings. In many cases, especially in pathological ones, the properties of the vocal folds often deviate from their generic values---sometimes asymmetrically wherein the characteristics of the two vocal folds differ for the same individual. In such cases, it is desirable to find a more scalable way to adjust the model parameters on a case by case basis. In this paper, we present a novel and alternate way to determine vocal fold model parameters from the speech signal. We focus on an asymmetric model and show that for such models, differences in estimated parameters can be successfully used to discriminate between voices that are characteristic of different underlying vocal fold pathologies.
\end{abstract}


\section{Introduction}
\label{sec:intro}
%The role of phonation in the bio-mechanical process of speech production has been widely studied in the literature. 
Phonation is the process wherein the vocal folds in the larynx are set into a state of self-sustained vibration, causing an excitation signal to be produced at the glottal source. This signal resonates in the vocal tract of the speaker and, depending on the shape of the vocal tract and the configuration of the articulators (tongue, lip, jaw, etc.), is heard as a characteristic voiced sound by the listener. Phonation is thus important in the production of all vowels and all voiced consonants in all languages of the world.

%The oscillation of the vocal folds is fundamental to phonation. During phonation, the airstream from lungs is modulated by the displacement of vocal folds, creating a glottal flow pattern~\cite{titze1988physics} that produces the fundamental frequency of voicing. The dynamic interaction of the glottal flow with the vocal tract generates different speech tones. 

As per the myoelastic-aerodynamic theory of phonation, the self-sustained vibrations of the vocal folds are initiated and driven by a delicate balance of physical and aerodynamic forces across the glottis
%, in which the muscles of the larynx and the dynamics of the airflow across it play a critical role
\cite{Cveticanin12}. The actual movements are governed by various biomechanical properties of the vocal folds such as elasticity, resistance, Young's modulus, viscosity, etc.
In pathological cases such as vocal palsy, phonotrauma, neoplasm, etc., these properties of the vocal structures vary from their generic settings ~\cite{bhat2018femh}. These often cause the movements of the vocal folds to become asymmetric~\cite{steinecke1995bifurcations}---where the movements of the left and right folds are out of sync---in a manner that is characteristic of the underlying pathology. 

The premise of our paper is that if these movements of the vocal folds and the underlying parameters of the system that produces them could be recovered from the speech signal, the underlying pathologies too could be identified\footnote{Note that this diverges from the traditional approach of classifying these through analysis of the surface-level waveform. However, it is not our intention in this paper to build a better mousetrap, but to present a new problem---that of uncovering vocal fold dynamics---that naturally leads to a new paradigm.}.
In order to do so, we must consider the actual {\em physics} of the vocal-fold movements, how they influence its movements, and how these manifest in the speech signal itself. 
%This is the problem addressed here.

The exact physics of the airflow through the glottis during phonation is well studied, e.g.,~\cite{flanagan1968self,ishizaka1972synthesis,titze1988physics,zhang2006influence,zhao2002computational,zhang2002computational}, and a number of physical models have been proposed for it, e.g.,~\cite{ishizaka1972synthesis, lucero1993dynamics, lucero2013modeling, alipour2000finite, yang2011computation, pickup2009influence, jiang01, singh2019profiling}.
%Each such model uses measured biomechanical properties of the vocal folds within a set of dynamically solvable equations that capture the balance of forces across the glottis during phonation. The output of each model is a trajectory in two or three-dimensional space that represents the self-sustained motion of the vocal folds.
The models use measured biomechanical properties of the vocal folds within a set of equations that capture the movements of the glottis during phonation. Since the vocal fold dynamics are non-linear, the models are systems of coupled non-linear dynamical equations as well.
%
%For a given set of parameters, their output is a trajectory that
%emulates the corresponding motions of the vocal folds. The
%trajectories themselves tend to fall into orbits that are characteristic of such systems -- with regular or irregular
%behaviors in phase space, the possible types, and distributions of which depend on the system characteristics.
%
%
They output a {\em phase space} trajectory of state variables that describes the movements of the folds.
The trajectories tend to fall into {\em orbits} with regular or irregular behaviors, which actually explain observed behavior patterns of the vocal folds.
The possible types and distributions of these orbits depend on the system parameters.
%
%The model output is a trajectory in two or three-dimensional space that represents the self-sustained motion of the vocal folds.
To explain {\em asymmetric} movements of vocal folds, such as those seen under pathologies, asymmetric dynamical models of the vocal folds have been proposed that are able to emulate these asymmetric vibratory motions of the vocal folds~\cite{erath2006investigation}.

While these models effectively solve the \textit{forward} problem of accurately emulating vocal fold movements during phonation, the \textit{reverse} problem of finding the correct model parameters given a set of observed vocal fold movements has not been addressed.
We develop a methodology to solve it based on the analysis of the speech signal. Given a model for asymmetric movements of the vocal folds and a set of speech signals from people affected by various pathologies that affect vocal fold movements, we propose a method to estimate the parameters of the asymmetric model that explains them.
We further show how the re-estimated model parameters can be mapped into the phase space of the nonlinear dynamical systems, and how the location of these parameters in the {\em parameter} space of the model can directly indicate the underlying pathology in the observed speech signal.

%The rest of this paper is organized as follows. In Section~\ref{sec:biomech}, we describe the biomechanical process of phonation and an asymmetrical physical model for it in detail. In Section~\ref{sec:phase}, we explain how the output of an asymmetrical model may be interpreted in both the phase space and variable space, and how both can be reconciled with speech signals. In Section~\ref{sec:method}, we show how model parameters may be re-estimated based on the speech signal. In Section~\ref{sec:expts} we present some observations and experimental results, and in Section~\ref{sec:concl} we present our conclusions.

\section {Phonation and the vocal folds model}\label{sec:biomech}
\begin{figure}[t!]
\centering
% \scalebox{0.8}{
    \includegraphics[width=0.9\linewidth]{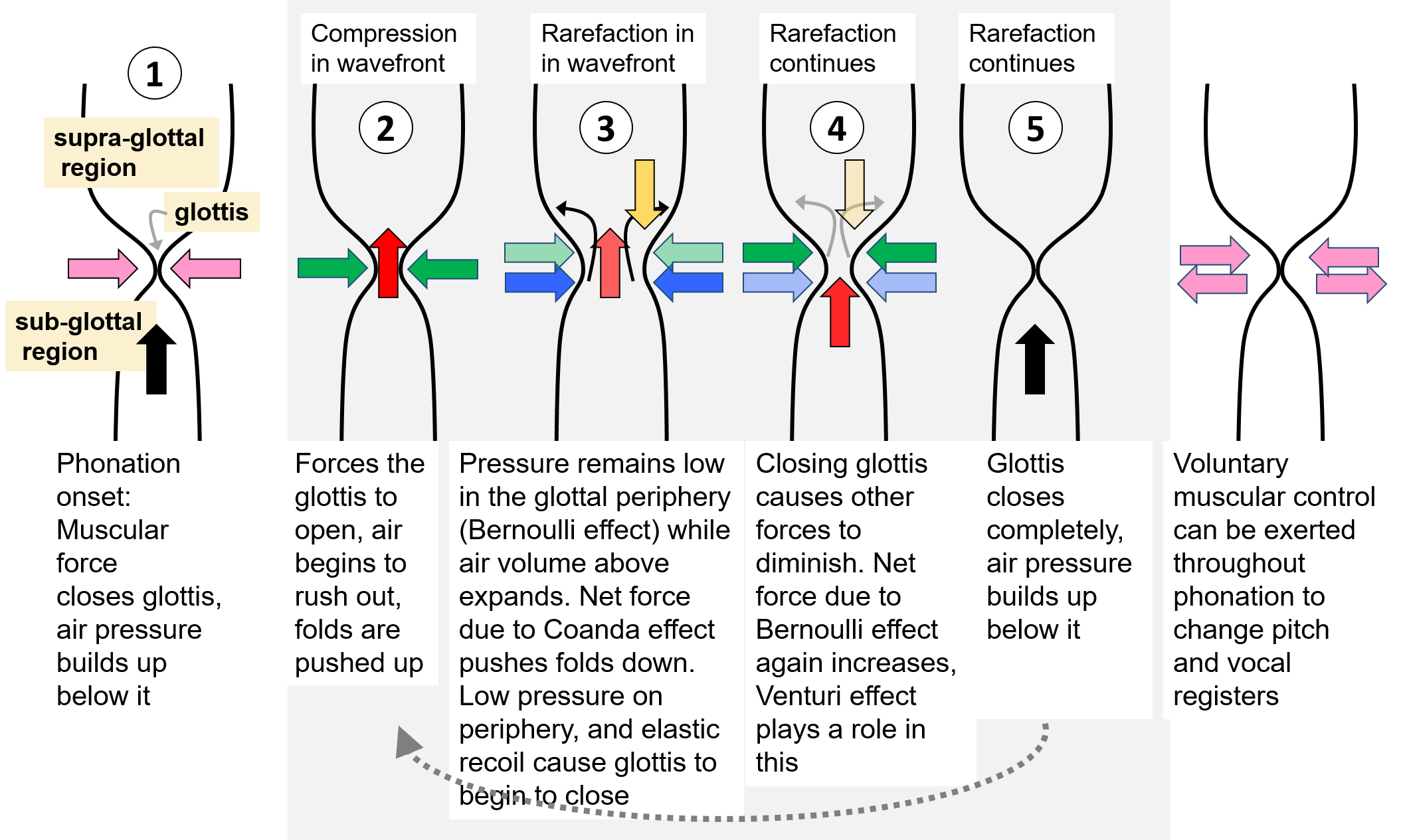}
% }
\caption{\footnotesize Schematic of the balance of forces through one cycle of the self-sustained vibrations of the vocal cords. The color codes for the arrows depict net forces due to the following: Pink--muscular; Green--Bernoulli effect; Yellow--Coand{\v a} effect; Blue--vocal fold elasticity and other factors; Black and Red--air pressure. Lighter shades of each color depict lower forces. Figure from~\cite{singh2019profiling} with permission.} 
\label{fig:coanda}
\end{figure}

At the biomechanical level, phonation happens as a result of a specific pattern of events in the glottal region.
The vocal folds are membranes that are set into vibratory motion as a result of a complex interplay of forces in the vocal tract. These relate to a) pressure balances and airflow dynamics within the supra-glottal and sub-glottal regions,
and b) muscular control within the glottis and the larynx. 
Specifically, during phonation, the vocal folds can vibrate hundreds of times per second. Such oscillations are self-sustained, a physical process driven by the right balance of opposing forces acting on it and set into motion as a result of a chain of events in the laryngeal region~\cite{titze1988physics}. The balance of forces necessary to cause self-sustained vibrations is created by two physical phenomena: the Bernoulli effect and the Coand{\v a} effect. Figure~\ref{fig:coanda} illustrates the interaction between these effects that drives the oscillations of the vocal folds.

\subsection{The asymmetric vocal folds oscillation model}
% \textit{\textbf{The asymmetric vocal folds oscillation model}}
To model phonation, computational models of vocal folds have been developed, including four broad types: $1$-mass models, e.g.~\cite{lucero2013modeling}, $2$-mass models, e.g.~\cite{ishizaka1972synthesis,lucero1993dynamics}, multi-mass models~\cite{yang2011computation}, and finite element models~\cite{alipour2000finite}. Each of these have proven to be useful in different contexts.
For the purpose of this study, the $1$-mass \textit{body-cover} model~\cite{titze1988physics,story1995voice,chan2006dependence,lucero2015self} is of particular interest.
It assumes that a glottal flow-induced \textit{mucosal wave} travels upwards within the transglottal region, causing a small displacement of the mucosal tissue, which attenuates down within a few millimeters into the tissue as an energy exchange happens between the airstream and the tissue~\cite{titze1988physics}.
This allows us to represent the mucosal wave as a one-dimensional surface wave on the mucosal surface (the cover), and treat the remainder of the vocal folds (the body) as a single mass or safely neglect it.
Under this assumption, the oscillation model can be linearized, and the oscillatory conditions are much simplified while maintaining the accuracy of the model.
% The vibration of vocal folds is modeled with a pair of mass-damper-spring oscillators.

In most vocal pathologies, it has been observed that the vocal fold vibrations are not symmetrical~\cite{pickup2009influence}. Thus the asymmetric vocal fold model is ideally suited to modeling the pathological phonation.
% Figure~\ref{fig:vocal_folds} shows a schematic diagram of the vocal folds and the $1$-mass body-cover model.
We adopt the specific formulation of the $1$-mass asymmetric model from~\cite{lucero2015self}.
This model incorporates an \textit{asymmetry parameter}, which describes the asymmetry in the oscillation of left and right vocal folds.
The key assumptions made in formulating this model are:
a) the degree of asymmetry is independent of the oscillation frequency,
b) the glottal flow is stationary, frictionless and incompressible,
% c) all subglottal and supraglottal loads are neglected, eliminating the effect of source-vocal tract interaction,
c) the effect of source-vocal tract interaction is eliminated,
d) there is no glottal closure and hence no vocal fold collision during the oscillation cycle,
and
% e) the one-dimensional wave that is assumed to travel within the mucosal surface has a small amplitude; thus, the body-cover model is employable.
e) the small-amplitude body-cover model is applicable.
% \begin{figure}[t]
%   \centering
% %   \scalebox{0.8}{
%   \includegraphics[width=0.45\linewidth]{images/vocal_folds.png}
% %   }
%   \caption{Diagram of vocal folds and $1$-mass body-cover model.
% %   The model assumes that the interaction between the airflow and the mucosal tissue at the glottis produces a (quasi-) periodic glottal flow, which acts as the source of phonation.
%   }
%   \label{fig:vocal_folds}
% \end{figure}

Following~\cite{lucero2015self} denote the center-line of the glottis as the $z$-axis.
At the midpoint ($z=0$) of the thickness of the vocal folds, the left and right vocal folds oscillate with lateral displacement $\xi_l$ and $\xi_r$, resulting in a pair of coupled Van der Pol oscillators
\begin{align}
    &\ddot{\xi}_r + \beta (1 + \xi_r^2)\dot{\xi}_r + \xi_r - \frac{\Delta}{2}\xi_r = \alpha (\dot{\xi}_r + \dot{\xi}_l)\nonumber\\
    &\ddot{\xi}_l + \beta (1 + \xi_l^2)\dot{\xi}_l + \xi_l + \frac{\Delta}{2}\xi_l = \alpha (\dot{\xi}_r + \dot{\xi}_l),\label{eq:mdl_vocal_fold}
\end{align}
where $\beta$ is the coefficient incorporating mass, spring and damping coefficients, $\alpha$ is the glottal pressure coupling coefficient, and $\Delta$ is the asymmetry coefficient.
For a male adult with normal voice, the reference values for the model parameters may be $\alpha=0.5$, $\beta=0.32$ and $\Delta=0$.

% \section {Phase space of the asymmetric model and its interpretation}\label{sec:phase}
\section {Phase space of the asymmetric model}\label{sec:phase}
To study the behaviors of nonlinear dynamical systems such as (\ref{eq:mdl_vocal_fold}), we introduce some useful concepts and tools.
The output variables of a dynamical system are also referred to as {\em state variables}, as they identify the current state of the system.
The \textbf{phase space} of a system is the space whose coordinates are its state variables. 
We can denote such system as $(T, M, \Phi)$, where $T=\sR_{\geq 0}$ is non-negative real time, the phase space $M$ is a differentiable manifold, and $\Phi: T \times M \supseteq U \to M$ is a continuous evolution function.
% As it evolves through time, the system charts a \textbf{trajectory} within its phase space. The behavior of the system is indicated by the shape and properties of this trajectory.
As the system evolves through time, the map $\Phi_x: T \to M$ is the \textbf{flow} or \textbf{trajectory} through $x \in M$, and the set of all flows $\gamma_x \coloneqq \{\Phi_x: t \in T \}$ is the \textbf{orbit} of $x$~\cite{birkhoff1927dynamical}.

% As a system evolves, its state trajectory can arrive at a \text{fixed point}, that does not change with further evolution, a \textbf{limit cycle}, where the state loops over a fixed number of values with a period, or even \textbf{chaos}, where the state stays within a localized region, but never repeats a value. Which of these behaviors a system exhibits depends on its parameters~\cite{jiang2001modeling,jiang2002chaotic}.

%An \textbf{attractor} is a closed region in the phase space of a system within which a system remains over time. Examples of attractors are \textbf{fixed points}, \textbf{limit cycles}, \textbf{strange attractors} etc. A strange attractor is a geometrical fractal structure that is the output of a system in chaos, within which no output is the same as another for infinite time. A fixed point is a point in the function's domain that is mapped to itself by the function. In an evolving dynamical system, a fixed point in phase space is defined by a set of variables or an output that does not vary with each time step. A limit cycle is a periodic orbit of a system in its phase space, within which the system remains over time.

The behaviors of flows can be described by attractors---a set $A \subseteq M$ that ``traps'' the orbit $\gamma_x$, i.e., starting from initial point $x$, there exists a $t_0$ such that $\Phi_x(t) \in A$ for $t > t_0$.
The simplest attractor is an equilibrium point.
But we are particularly interested in those attractors revealing the periodic motion of the flow in phase space.
Such attractors include the \textbf{limit cycle} or the \textbf{limit torus}, which is an isolated periodic or toroidal orbit.
Some other attractors are chaotic, whose trajectories diverge with arbitrarily close initial positions~\cite{jiang2001modeling,jiang2002chaotic}. The phase space of the system may include multiple attractors, the number, location, and type of which depend on its parameters.

% A y given setti\textbf{bifurcation diagram} shows the change in the system's limiting behavior as a function of its parameters. Typically, as parameters vary, the system changes from converging to fixed points, to limit cycles of increasing period, and eventually to chaos.
Periodic attractors reveal the stability of the system.
To dissect the structure of attractors, we use the \textbf{Poincar\'e map}.
For an $n$-dimensional phase space with a periodic orbit $\gamma_x$, a Poincar\'e section $S$ is an $(n-1)$-dimensional section (hyper-plane) transversal to $\gamma_x$, and the Poincar\'e map on $S$ is a map $P: S \supseteq U \to S$, $x \mapsto \Phi_x(t_s)$, where $t_s$ is the time between two intersections~\cite{birkhoff1927dynamical}.
We also want to study how the topological structure, including attractors, of phase space change with system parameters.
This leads to the study of \textbf{bifurcation}, which occurs when a small change of a model parameter causes an abrupt change of the topological structure. A {\em bifurcation diagram} is a visualization of the parameter space of the system showing the number (and behavior) of attractors at each parameter setting.
%For each parameter setting, given the initial conditions of the displacement of vocal folds, such diagrams can provide an indication of what the glottal flow pattern may be at a future time, e.g., \cite{berry1994interpretation}
%one that indicates the behavior of a system with each setting of its parameter values. Thus the co-ordinates of such a diagram are the parameters (or a subset of the parameters) of the system.

% While the behavior and phase trajectory of a system may be determined from knowledge of the dynamical system and its parameters, the premise of this paper is that the inverse can also be performed---the parameters of the system may be inferred from its phase trajectory. In the case of vocal fold dynamics, this inversion recovers indelible cues to underlying vocal pathologies. 
%\subsection{Voice disorders and the significance of the asymmetric model}
\subsection{Physical interpretation of phase space of asymmetric model}

Back to our context, the dynamical system of concern is the asymmetric vocal folds model (\ref{eq:mdl_vocal_fold}).
Its phase space, which is four-dimensional, includes states $(\xi_r, \d{\xi}_r, \xi_l, \d{\xi}_l)$.
For this nonlinear system, it is expected that attractors such as limit cycles or toruses will appear in the phase space.
Such phenomena are consequences of specific parameter settings.
Specifically, the parameter $\beta$ determines the periodicity of oscillations; the parameter $\alpha$ and $\Delta$ quantify the asymmetry of the displacement of left and right vocal folds and the degree to which one of the vocal folds is out of phase with the other~\cite{steinecke1995bifurcations,lucero2015self}.
We can visualize them by plotting the left and right displacements as well as the phase space portrait.

\begin{figure}[t!]
  \centering
  \includegraphics[width=0.9\linewidth]{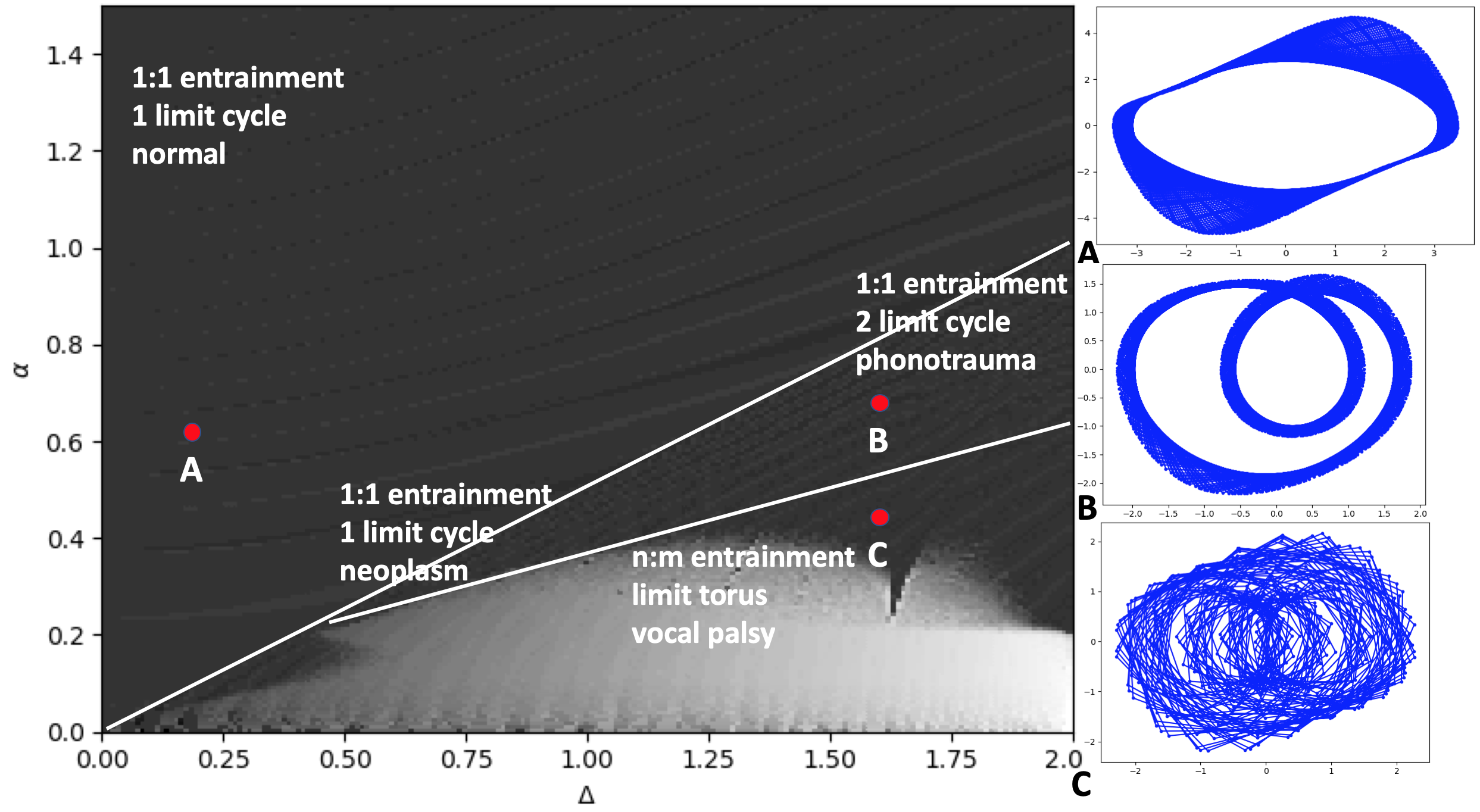}
  \caption{\footnotesize Bifurcation diagram of the asymmetric vocal fold model. It shows the entrainment ratio $n : m$ (coded as different shades) as a function of model parameter $\alpha$ and $\Delta$, where $n$ and $m$ are the number of intersections of the orbits of right and left oscillators across the Poincar\'e section $\d{\xi}_{r,l}=0$ at stable status. This is consistent with the theoretical results in~\cite{lucero2015self}.
  The phase portraits for point A, B, C are shown alongside, where the horizontal axis is displacement, and the vertical axis is velocity.}
  \label{fig:bifurcation_plots}
\end{figure}

The coupling of right and left oscillators is described by their \textbf{entrainment}; they are in $n:m$ entrainment if their phase $\theta_r$, $\theta_l$ satisfy $|n\theta_r - m\theta_l| < C$ where $n, m$ are integers and $C$ is a constant~\cite{lucero2015self}.
Such entrainment can be shown by the Poincar\'e map, where the number of crossings of the trajectory of right or left oscillator with the Poincar\'e section shows the periodicity of its limit cycles. Therefore, their ratio represents the entrainment.
To show how the entrainment changes with parameters, we plot the bifurcation diagram.
An example of such a bifurcation diagram is shown in Figure~\ref{fig:bifurcation_plots}~\cite{steinecke1995bifurcations,lucero1993dynamics}.
% which depicts the behavior of the dynamical system for the vocal folds against initial parameter settings of $\alpha$ (the glottal pressure coupling coefficient) and $\Delta$ (the asymmetry coefficient).
As we will see later (and as indicated in Figure \ref{fig:bifurcation_plots}), model parameters can characterize voice pathologies, which will also be visible in phase portrait and bifurcation plots.
% Such asymmetries are characteristic of various voice disorders and differ depending on the type of voice disorder. This also means that if we were able to determine the degree of asymmetry in this model or the parameter $\Delta$, given a speech signal, we would be able to identify the type of underlying voice pathology that may have caused it.

\section {Parameter estimation for the model}\label{sec:method}
% It is first important to understand why we may want to estimate the parameters of the model in Equation~(\ref{eq:mdl_vocal_fold}) based on speech. This has a powerful incentive as described below.

% \subsection{Model parameters estimation}
Finding the parameters of any physical model that emulates vocal fold oscillations is not trivial.
For this, one must acquire measurements of the vocal fold displacements, which in turn require either high-speed photography~\cite{mergell2000irregular} or physical or numerical simulations~\cite{alipour2000finite,tao2007asymmetric}, which are often not easily accessible.
{\em Even} with the measurements, estimating the model parameters remains hard. The problem itself is commonly termed as the \emph{inverse problem}~\cite{isakov2006inverse}, and is usually solved via iterative matching procedures~\cite{tao2004estimating,zhang2006parameter,rupitsch2011simulation}, stochastic optimization or heuristic procedures~\cite{yang2011computation,tao2007extracting}. 

We propose a method for solving the inverse problem that bypasses the difficulties inherent in traditional methods: namely that of either obtaining direct measurements of vocal fold displacements or of the complexity of solving inverse problems using traditional methods.
Our proposed solution comprises an \textbf{Adjoint Least-Squares method}, which we call \textbf{ADLES} for brevity, to estimate model parameters directly from speech measurements.
% Based on this, we also propose model-state-based metrics to quantify and analyze the irregular oscillatory modes of the vocal folds that characterize different types of voice disorders.

% \subsection{Parameter Estimation}
First, we formulate our objective.
The vibration of vocal folds oscillates the air particles at the glottal region, producing a pressure wave that propagates through the upper vocal channel into the open air.
This pressure wave is considered planar when its frequency is under $4$ KHz~\cite{sondhi1974model}, and hence a function of position $x \in \varOmega \coloneqq [0, L]$ and time $t \in \gT \coloneqq [0, T]$---$p(x,t) \in \gL^2(\varOmega \times \gT)$, where $L$ is the length of the upper vocal channel.
The acoustic pressure $p_L(t) \coloneqq p(L, t)$, which represents the speech signal measured by a microphone near the mouth, is a result of the pressure wave $p_0(t) \coloneqq p(0, t)$ at the glottis modulated by the upper vocal channel.
If we denote the effect of upper vocal channel as a filter 
\begin{align}
\gF: & \gL^2(T) \to \gL^2(T)\\
     & p_0(t) \mapsto p_L(t),
\label{eq:filter_operator}
\end{align}
we can deduce $p_0(t)$ from $p_L(t)$ using inverse filtering~\cite{alku2011glottal}
\begin{equation}
    p_0(t) = \gF^{-1}(p_L(t)).
\label{eq:p0_inv_filt}
\end{equation}
Let $A(x)$ be the area function of the vocal channel for $x \in [0, L]$ and $A(0)$ represent the cross-sectional area at the glottis. The corresponding volume velocity $u_0(t)$ through the vocal channel is given by 
\begin{equation}
    u_0(t) = \frac{A(0)}{\rho c}p_0(t),
\label{eq:u0_p0}
\end{equation}
where $c$ is the speed of sound and $\rho$ is the ambient air density.
As a result, given a measured speech signal $p_m(t)$, we have
\begin{equation}
    u_0^m(t) = \frac{A(0)}{\rho c}\gF^{-1}(p_m(t)).
\label{eq:u0_m_F_inv}
\end{equation}
Alternatively, we can derive $u_0(t)$ from the displacement of vocal folds by
\begin{equation}
    u_0(t) = \t{c}d\L(2\xi_0 + \xi_l(t) + \xi_r(t)\R),
\label{eq:u0_xi}
\end{equation}
where $\xi_0$ is the half glottal width at rest and is set to $0.1$ cm, $d$ is the length of vocal fold and is set to $1.75$ cm, and $\t{c}$ is the air particle velocity at the midpoint of the vocal fold.
Our objective is then to minimize the difference
\begin{align}
    & \min \int_{0}^{T} \L(u_0(t) - u_0^m(t)\R)^2 dt \Leftrightarrow \\
	& \min \int_{0}^{T} \L(\t{c}d\L(2\xi_0 + \xi_l(t) + \xi_r(t)\R) - \frac{A(0)}{\rho c}\gF^{-1}(p_m(t))\R)^2 dt,
\label{eq:obj_least_squares}
\end{align}
such that
\begin{align}
\quad & \ddot{\xi}_r + \beta (1 + \xi_r^2)\dot{\xi}_r + \xi_r - \frac{\Delta}{2}\xi_r = \alpha (\dot{\xi}_r + \dot{\xi}_l) \label{eq:x_r}\\
      &\ddot{\xi}_l + \beta (1 + \xi_l^2)\dot{\xi}_l + \xi_l + \frac{\Delta}{2}\xi_l = \alpha (\dot{\xi}_r + \dot{\xi}_l) \label{eq:x_l}\\
      &\xi_r(0) = C_r \label{eq:xr_init} \\
      &\xi_l(0) = C_l \label{eq:xl_init} \\
      &\d{\xi}_r(0) = 0 \label{eq:dxr_init} \\
      &\d{\xi}_l(0) = 0, \label{eq:dxl_init}
\end{align}
where $C_r$ and $C_l$ are constants.

\subsection{The Adjoint Least Squares solution}
To solve the functional least squares in (\ref{eq:obj_least_squares}), we require the gradients of (\ref{eq:obj_least_squares}) w.r.t. the model parameters $\alpha$, $\beta$ and $\Delta$. Subsequently we can adopt any gradient based (local or global) method to obtain the solution.
Denote the residual $R = \t{c}d\L(2\xi_0 + \xi_l(t) + \xi_r(t)\R) - \frac{A(0)}{\rho c}\gF^{-1}(p_m(t))$;
then $f(\xi_l, \xi_r; \vartheta) = R^2$, and $F(\xi_l, \xi_r; \vartheta) = \int_0^T f(\xi_l, \xi_r; \vartheta) dt$, where $\vartheta = [\alpha, \beta, \Delta]$.
We define the Lagrangian
\begin{align}
    \mathscr{L}(\vartheta) &= \int_0^T \L[ f + \lambda \L( \ddot{\xi}_r + \beta \L(1 + \xi_r^2\R)\dot{\xi}_r + \xi_r - \frac{\Delta}{2}\xi_r - \alpha \L(\dot{\xi}_r + \dot{\xi}_l\R) \R) \R. \nonumber\\
    & + \L. \eta \L( \ddot{\xi}_l + \beta \L(1 + \xi_l^2\R)\dot{\xi}_l + \xi_l + \frac{\Delta}{2}\xi_l - \alpha \L(\dot{\xi}_r + \dot{\xi}_l \R) \R) \R] dt \nonumber\\
    & + \mu_l \L(\xi_l(0) - C_l\R) + \mu_r \L(\xi_r(0) - C_r\R) + \nu_l \d{\xi}_l(0) + \nu_r \d{\xi_r}(0),
\label{eq:lagrangian}
\end{align}
where $\lambda$, $\eta$, $\mu$ and $\nu$ are Lagrangian multipliers.
Taking the derivative of the Lagrangian w.r.t. the model parameter $\alpha$ yields
\begin{align}
    \mathscr{L}_{\alpha} &= \int_0^T \L[ 2\t{c}d R (\p_{\alpha}\xi_l + \p_{\alpha}\xi_r ) \R. \nonumber\\
    & + \lambda \L( \p_{\alpha}\ddot{\xi}_r + 2\beta\d{\xi}_r\xi_r\p_{\alpha}\xi_r + \beta \L(1 + \xi_r^2\R)\p_{\alpha}\d{\xi}_r \R. \nonumber\\
	& + \L. \p_{\alpha}\xi_r - \frac{\Delta}{2}\p_{\alpha}\xi_r - \alpha \L(\p_{\alpha}\d{\xi}_r + \p_{\alpha}\d{\xi}_l\R) - \L(\d{\xi}_r + \d{\xi}_r\R) \R) \nonumber\\
    & + \eta \L( \p_{\alpha}\ddot{\xi}_l + 2\beta\d{\xi}_l\xi_l\p_{\alpha}\xi_l + \beta \L(1 + \xi_l^2\R)\p_{\alpha}\d{\xi}_l \R. \nonumber\\
	& + \L. \L. \p_{\alpha}\xi_l + \frac{\Delta}{2}\p_{\alpha}\xi_l - \alpha \L(\p_{\alpha}\d{\xi}_r + \p_{\alpha}\d{\xi}_l\R) - \L(\d{\xi}_r + \d{\xi}_r\R) \R) \R] dt \nonumber\\
    & + \mu_l \p_{\alpha}\xi_l(0) + \mu_r \p_{\alpha}\xi_r(0) + \nu_l \p_{\alpha}\d{\xi}_l(0) + \nu_r \p_{\alpha}\d{\xi_r}(0).
\label{eq:L_a}
\end{align}
Integrating the term $\lambda \p_{\alpha}\ddot{\xi}_r$ twice by parts yields
\begin{equation}
    \int_0^T \lambda \p_{\alpha}\ddot{\xi}_r dt = \int_0^T \p_{\alpha}\xi_r \ddot{\lambda} dt  - \p_{\alpha}\xi_r\d{\lambda} \rvert_0^T + \p_{\alpha}\d{\xi}_r \lambda \rvert_0^T.
\label{eq:int_parts_lmd}
\end{equation}
Applying the same to $\eta \p_{\alpha}\ddot{\xi}_l$, substituting into (\ref{eq:L_a}) and simplifying the final expression we obtain
\begin{align}
    \mathscr{L}_{\alpha} &= \int_0^T \L[ \L(\ddot{\lambda} + \L(2\beta\xi_r\d{\xi}_r + 1 - \frac{\Delta}{2}\R)\lambda + 2\t{c}d R \R)\p_{\alpha}\xi_r \R. \nonumber\\
	& + \L(\ddot{\eta} + \L(2\beta\xi_l\d{\xi}_l + 1 + \frac{\Delta}{2}\R)\lambda +  2\t{c}d R \R)\p_{\alpha}\xi_l \nonumber\\
    & + \L( \beta(1 + \xi_r^2)\lambda - \alpha (\lambda + \eta) \R) \p_{\alpha} \d{\xi}_r \nonumber\\
	& + \L( (\beta(1 + \xi_l^2)\eta - \alpha(\lambda + \eta) \R) \p_{\alpha} \d{\xi}_l \nonumber\\
	& \L. - (\d{\xi}_r + \d{\xi}_l)(\lambda + \eta) \R] dt \nonumber\\
    & + \L(\mu_r + \d{\lambda}\R)\p_{\alpha}\xi_r(0) - \d{\lambda}\p_{\alpha}\xi_r(T) \nonumber\\
	& + \L(\nu_r - \lambda\R)\p_{\alpha}\d{\xi}_r(0) + \lambda \p_{\alpha}\d{\xi}_r(T) \nonumber\\
    & + \L(\mu_l + \d{\eta}\R)\p_{\alpha}\xi_l(0) - \d{\eta}\p_{\alpha}\xi_l(T) \nonumber\\
	& + \L(\nu_l - \eta\R)\p_{\alpha}\d{\xi}_l(0) + \eta\p_{\alpha}\d{\xi}_l(T).
\label{eq:L_a_simplify}
\end{align}
Since the partial derivative of the model output $\xi$ w.r.t. the model parameter $\alpha$ is difficult to compute, we eliminate the related terms by setting
\begin{align}
    &\mathrm{For}\; 0 < t < T: \nonumber\\
    &\quad \ddot{\lambda} + \L(2\beta\xi_r\d{\xi}_r + 1 - \frac{\Delta}{2}\R)\lambda + 2\t{c}d R = 0 \label{eq:adj_L}\\
    &\quad \ddot{\eta} + \L(2\beta\xi_l\d{\xi}_l + 1 + \frac{\Delta}{2}\R)\eta + 2\t{c}d R = 0 \label{eq:adj_E} \\
    &\quad \beta\L(1 + \xi_r^2\R)\lambda - \alpha \L(\lambda + \eta\R) = 0 \label{eq:adj_alg_1}\\
    &\quad \beta\L(1 + \xi_l^2\R)\eta - \alpha \L(\lambda + \eta\R) = 0, \label{eq:adj_alg_2}
\end{align}
with initial conditions
\begin{align}		
    &\mathrm{At}\; t = T: \nonumber\\
    &\quad \lambda(T) = 0 \label{eq:L_init}\\
    &\quad \d{\lambda}(T) = 0 \label{eq:dL_init}\\
    &\quad \eta(T) = 0 \label{eq:E_init}\\
    &\quad \d{\eta}(T) = 0. \label{eq:dE_init}
\end{align}
As a result, we obtain the derivative of $F$ w.r.t. $\alpha$ as
\begin{equation}
    F_{\alpha} = \int_0^T - \L(\d{\xi}_r + \d{\xi}_l\R)(\lambda + \eta) dt.
\label{eq:F_a}
\end{equation}
The derivatives of $F$ w.r.t. $\beta$ and $\Delta$ are similarly obtained as
\begin{align}
F_{\beta} &= \int_0^T \L( \L(1 + \xi_r^2\R)\d{\xi}_r\lambda + \L(1 + \xi_l^2\R)\d{\xi}_l\eta \R) dt \label{eq:F_b}\\
F_{\Delta} &= \int_0^T \tfrac1{2}\L(\xi_l\eta - \xi_r\lambda\R) dt. \label{eq:F_d}
\end{align}
Having calculated the gradients of $F$ w.r.t. the model parameters, we can now apply gradient-based algorithms to optimize our objective (\ref{eq:obj_least_squares}). 
For instance, applying gradient descent, we have
\begin{align}
    \alpha^{k+1} &= \alpha^{k} - \tau^{\alpha} F_{\alpha} \nonumber\\%\label{eq:grad_descent_a}
    \beta^{k+1} &= \beta^{k} - \tau^{\beta} F_{\beta} \nonumber\\%\label{eq:grad_descent_b}
    \Delta^{k+1} &= \Delta^{k} - \tau^{\Delta} F_{\Delta}, %\label{eq:grad_descent_d}
    \label{eq:grad_descent_updates}
\end{align}
where $\tau^{\cdot}$ is the step-size.
The overall algorithm is summarized as follows:
\begin{enumerate}
    \item Integrate (\ref{eq:x_r}) and (\ref{eq:x_l}) with initial conditions (\ref{eq:xr_init}), (\ref{eq:xl_init}), (\ref{eq:dxr_init}) and (\ref{eq:dxl_init}) from $0$ to $T$, obtaining $\xi_r$, $\xi_l$, $\d{\xi}_r$, and $\d{\xi}_l$.
    \item Solve (\ref{eq:adj_L}), (\ref{eq:adj_E}), (\ref{eq:adj_alg_1}) and (\ref{eq:adj_alg_2}) with initial conditions (\ref{eq:L_init}), (\ref{eq:dL_init}), (\ref{eq:E_init}) and (\ref{eq:dE_init}) from $T$ to $0$, obtaining $\lambda$, $\d{\lambda}$, $\eta$, and $\d{\eta}$.
    \item Update $\alpha$, $\beta$, and $\Delta$ with (\ref{eq:grad_descent_updates}).
\end{enumerate}
%\subsection{Convergence and Stability Analysis}
%Existence of solution
%Convergence of algorithm
%Sensitivity analysis
%Stability analysis

\section {Experiments}\label{sec:expts}
% Voice disorders refer to the abnormality when voice quality differs from its normal status~\cite{titze1998principles}. The abnormality can be physiological, i.e., due to the structural alteration of voice apparatus such as edema or vocal nodules, or due to neurogenic changes such as vocal tremor, spasmodic dysphonia or paralysis of vocal folds. The abnormality can also be functional, i.e., due to the improper use of voice apparatus such as vocal fatigue, muscle tension dysphonia, aphonia, diplophonia, or ventricular phonation.

In our experiments, we show the validity of our proposed ADLES method by using it to estimate the asymmetric model parameters for clinically acquired pathological speech data.
We show that the estimated parameters can then be effectively used to characterize the vocal disorders represented in our experimental data.

The dataset used in our experiments is the FEMH dataset~\cite{bhat2018femh}. It has $200$ voice samples of sustained vowel sound /a:/ obtained from a voice clinic in a tertiary teaching hospital, which includes $50$ normal voice samples and $150$ samples of common voice disorders. Within the disordered samples, there are $40/60/50$ samples for glottis neoplasm, phonotrauma (including vocal nodules, polyps, and cysts), and unilateral vocal paralysis, respectively.

\begin{figure}[t!]
\centering
\includegraphics[width=0.45\textwidth]{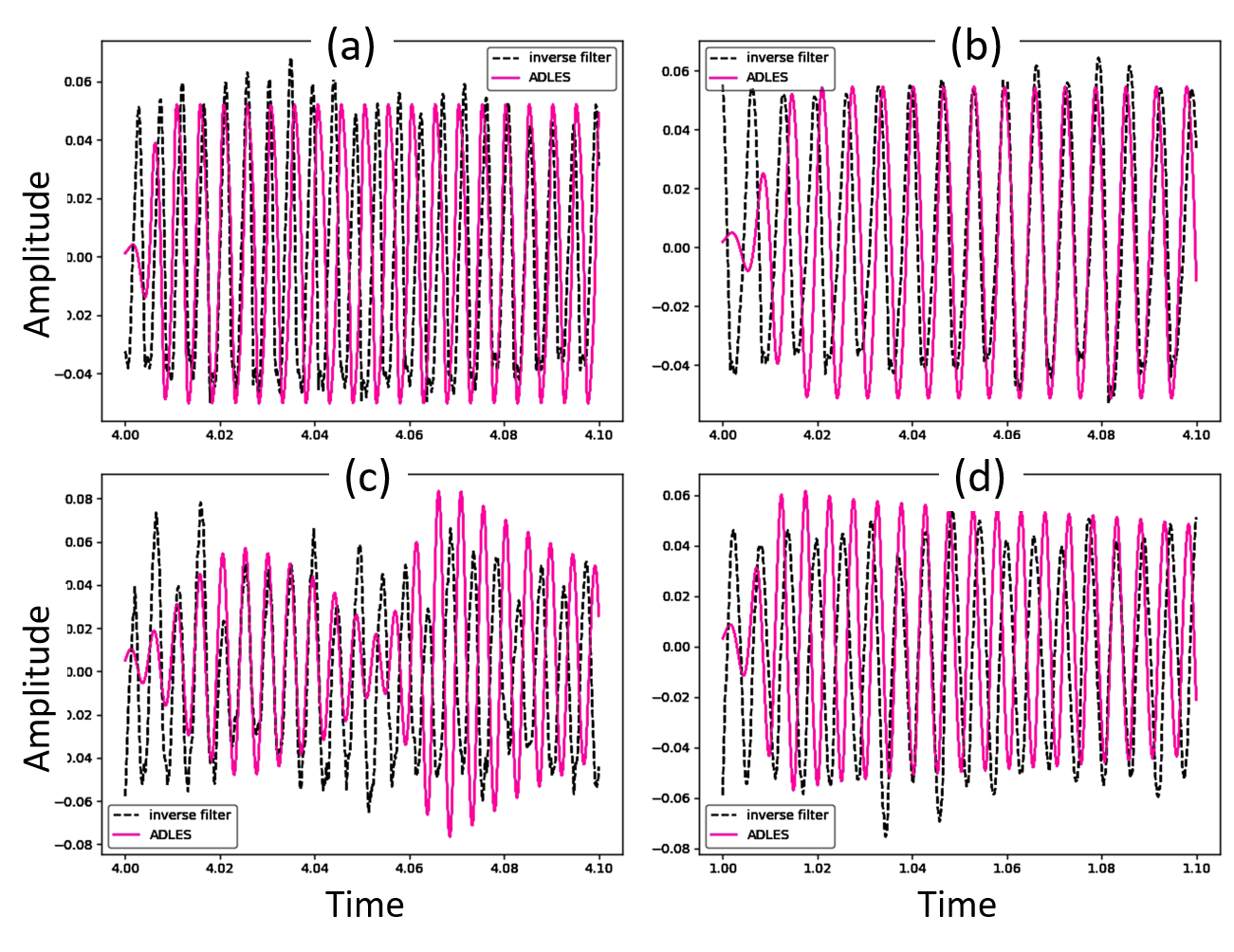}
    % \begin{subfigure}[b]{0.4\columnwidth}
    %     % \includegraphics[width=\textwidth]{images/glottal_flow_normal.png}
    %     \includegraphics[width=\textwidth]{images/glottal_flow_normal_new1.png}
    %     \caption{}
    %     \label{sfig:glottal_flow_normal}
    % \end{subfigure}
    % ~
    % \begin{subfigure}[b]{0.4\columnwidth}
    %     % \includegraphics[width=\textwidth]{images/glottal_flow_neoplasm.png}
    %     \includegraphics[width=\textwidth]{images/glottal_flow_neoplasm_new1.png}
    %     \caption{}
    %     \label{sfig:glottal_flow_neoplasm}
    % \end{subfigure}
    % \hfill
    % \begin{subfigure}[b]{0.4\columnwidth}
    %     % \includegraphics[width=\textwidth]{images/glottal_flow_phonotrauma.png}
    %     \includegraphics[width=\textwidth]{images/glottal_flow_phonotrauma_new1.png}
    %     \caption{}
    %     \label{sfig:glottal_flow_phonotrauma}
    % \end{subfigure}
    % ~
    % \begin{subfigure}[b]{0.4\columnwidth}
    %     % \includegraphics[width=\textwidth]{images/glottal_flow_vocalpalsy.png}
    %     \includegraphics[width=\textwidth]{images/glottal_flow_vocalpalsy_new1.png}
    %     \caption{}
    %     \label{sfig:glottal_flow_vocalpalsy}
    % \end{subfigure}
    \vspace{-0.05in}
    \caption{\footnotesize Glottal flows from inverse filtering and our ADLES estimation for (a) normal speech, (b) neoplasm, (c) phonotrauma, (d) vocal palsy.}
    \label{fig:glottal_flow_plots}
\end{figure}

\begin{figure}[h!]
    \centering
    \begin{subfigure}[b]{0.4\columnwidth}
        \includegraphics[width=\textwidth]{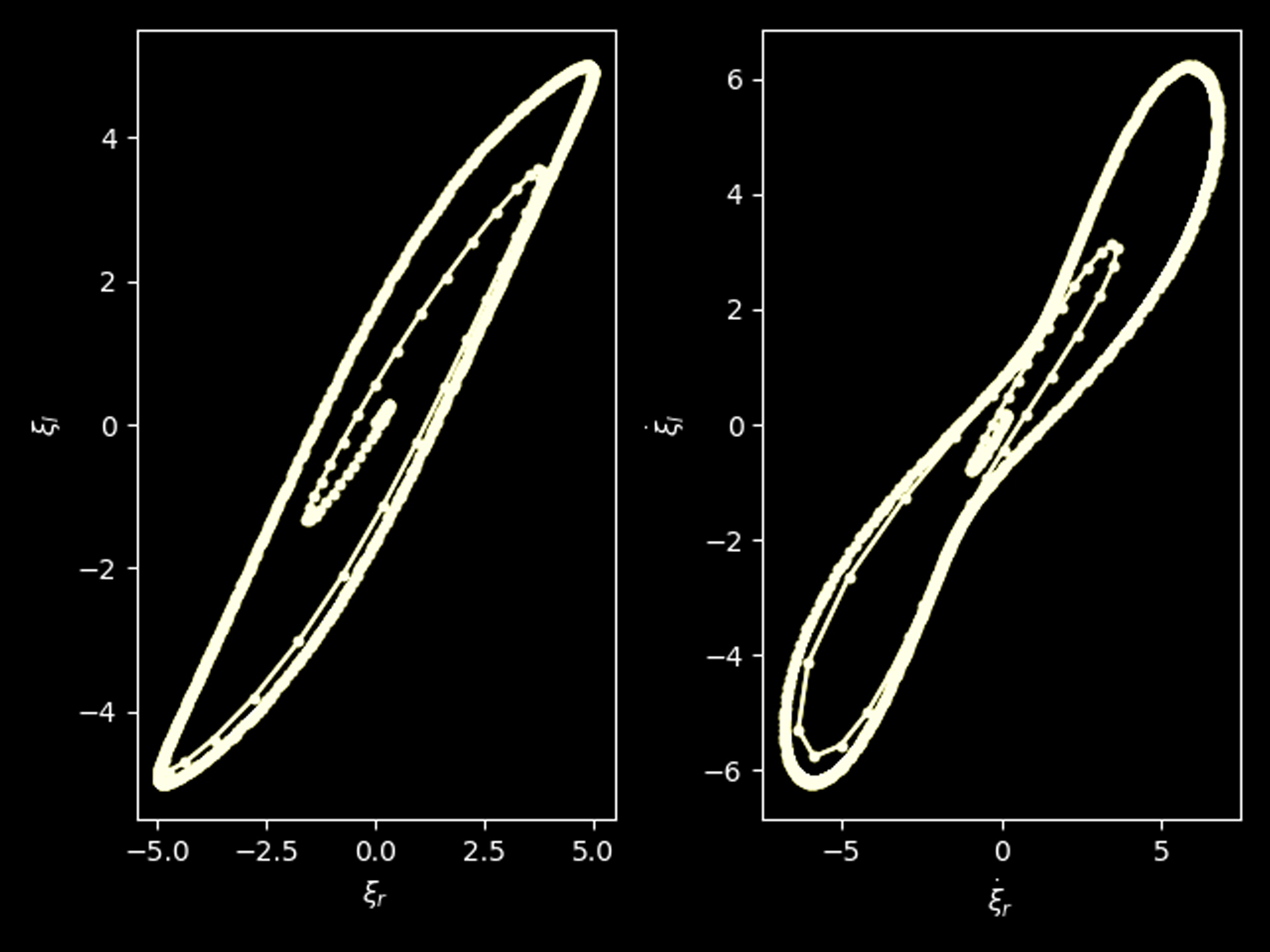}
        \caption{}
        \label{sfig:phase_plot_normal}
    \end{subfigure}
    ~
    \begin{subfigure}[b]{0.4\columnwidth}
        \includegraphics[width=\textwidth]{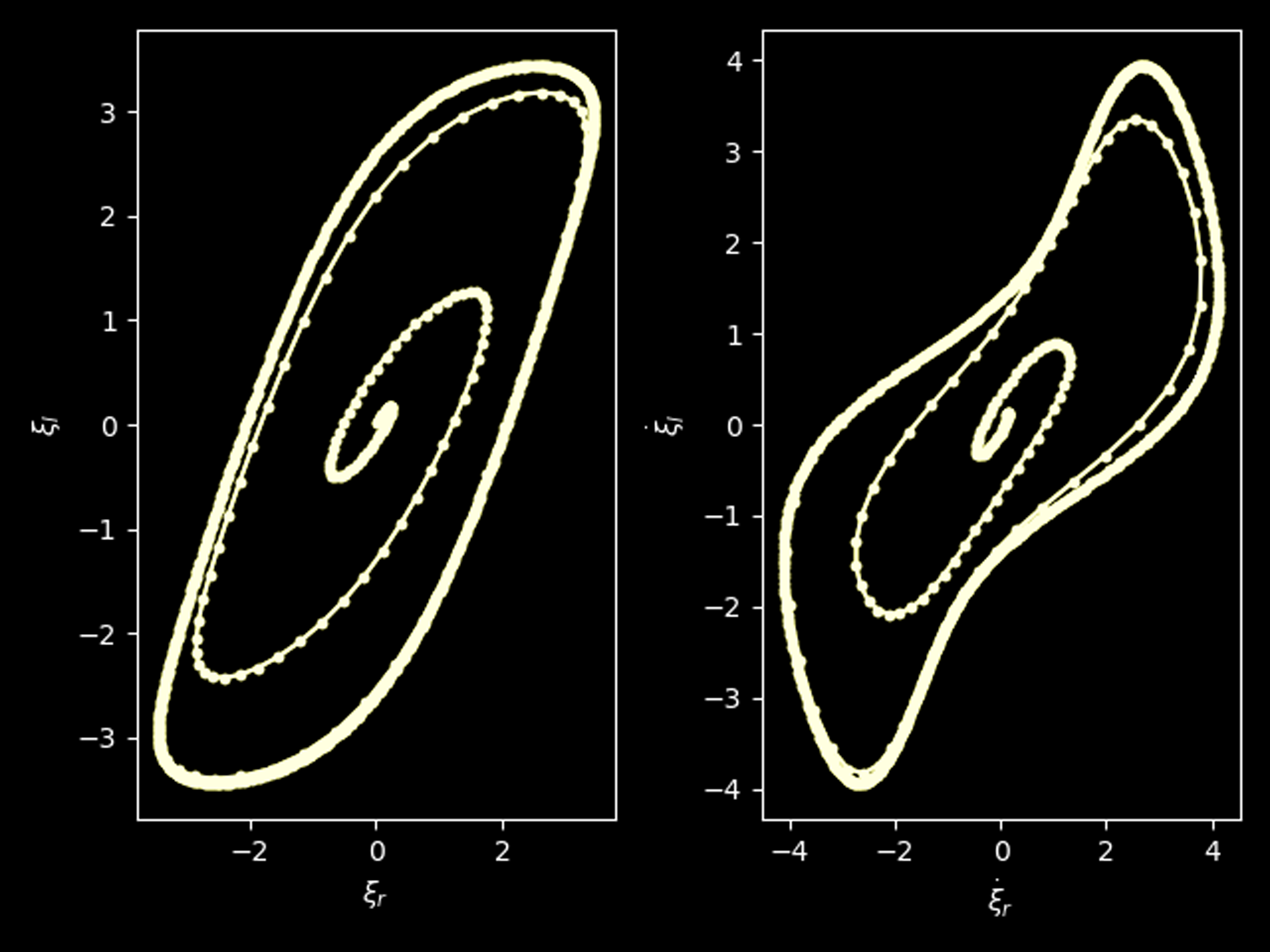}
        \caption{}
        \label{sfig:phase_plot_neoplasm}
    \end{subfigure}
    \hfill
    \begin{subfigure}[b]{0.4\columnwidth}
        \includegraphics[width=\textwidth]{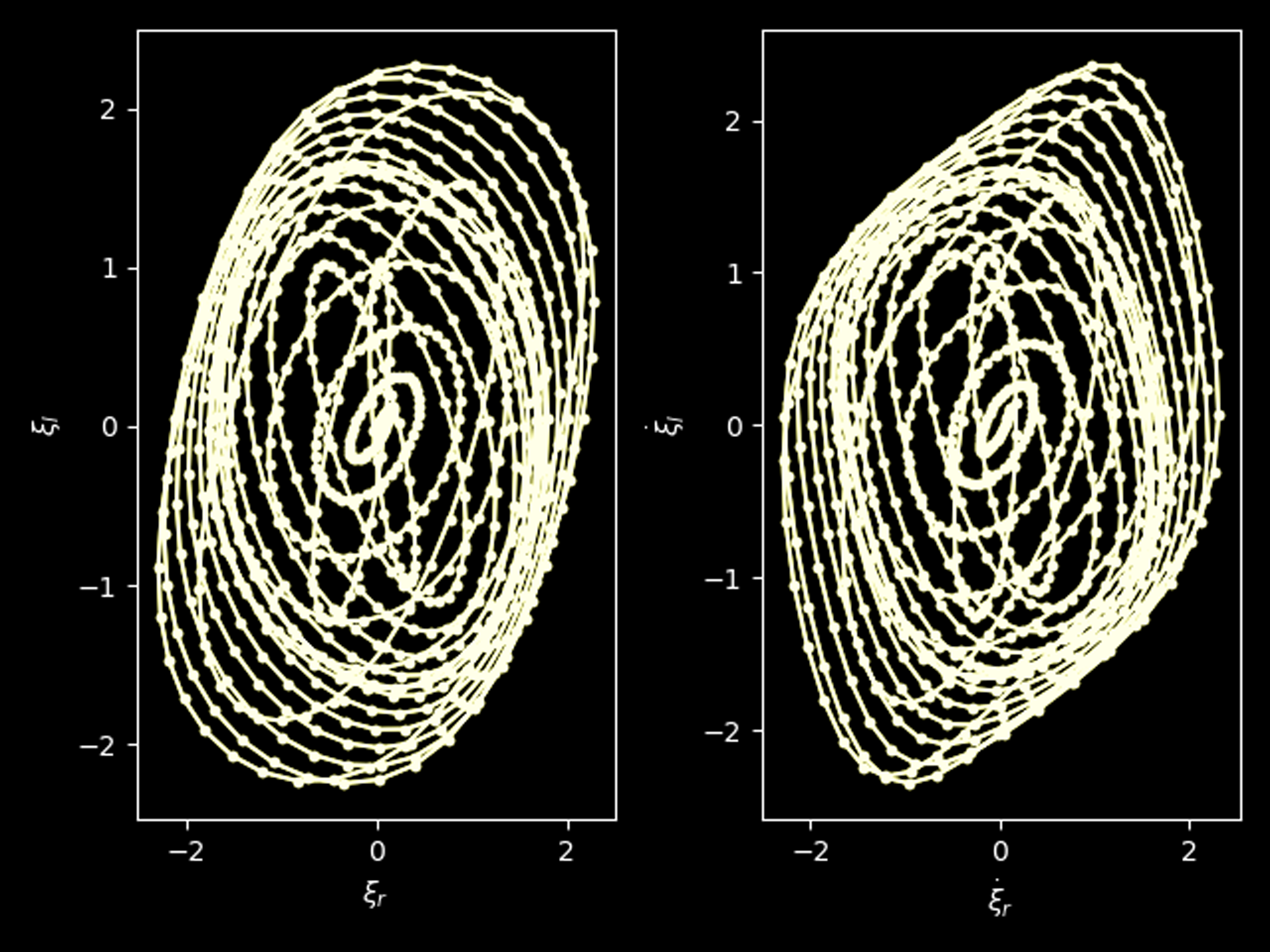}
        \caption{}
        \label{sfig:phase_plot_phonotrauma}
    \end{subfigure}
    ~
    \begin{subfigure}[b]{0.4\columnwidth}
        \includegraphics[width=\textwidth]{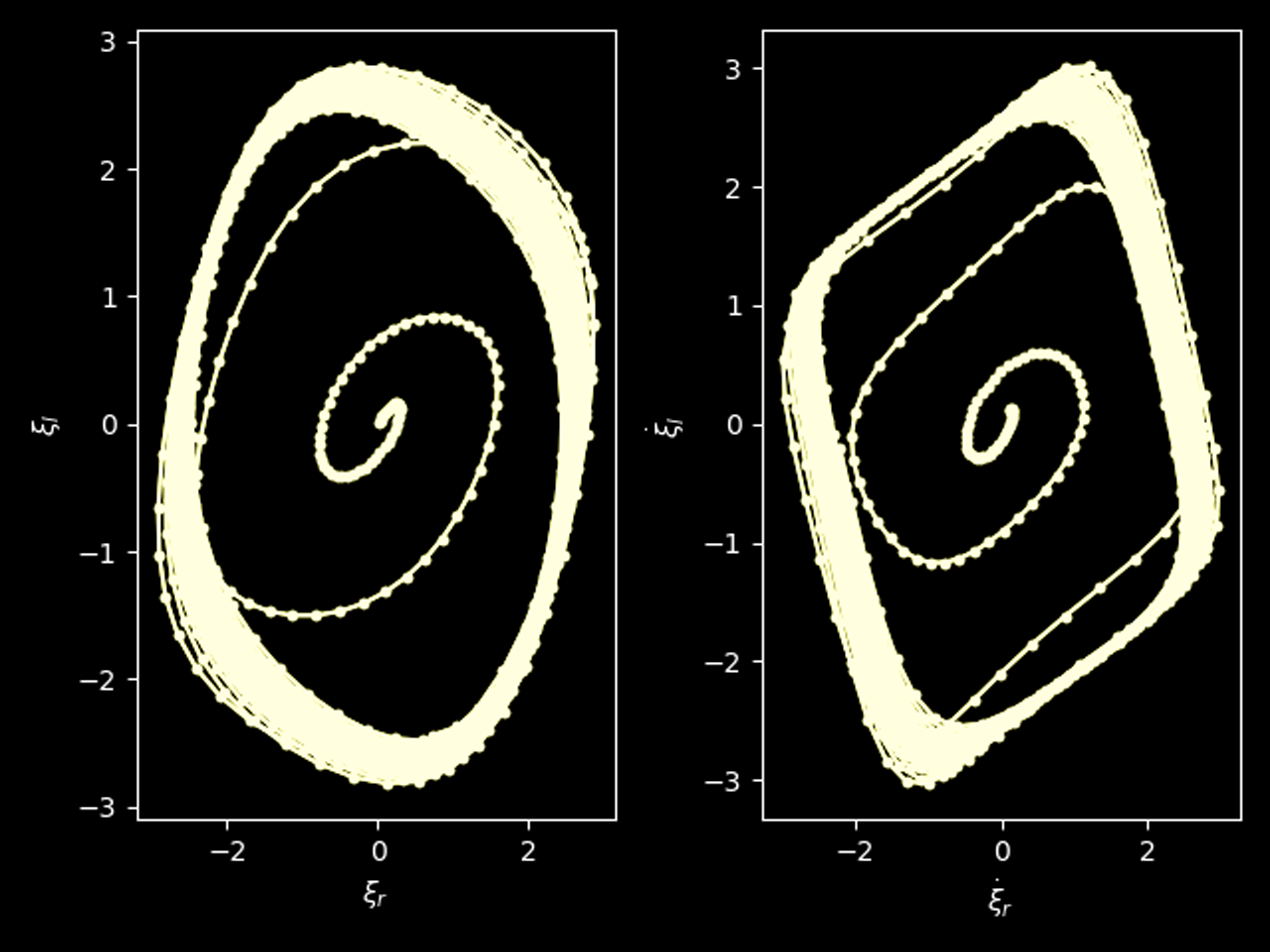}
        \caption{}
        \label{sfig:phase_plot_vocalpalsy}
    \end{subfigure}
    \vspace{-0.1in}
    \caption{\footnotesize Phase portraits of the system from our ADLES estimation for (a) normal speech--1 limit cycle, (b) neoplasm--1 limit cycle, (c) phonotrauma--2 limit cycles, (d) vocal palsy--limit torus. In each panel the left plot shows $\xi_l$ vs $\xi_r$ and the right plot shows $\dot{\xi}_l$ vs $\dot{\xi}_r$.} 
    \label{fig:phase_plots}
\end{figure}

% \subsection{Simulation results}
% \textbf{\textit{Results}}
Figure~\ref{fig:glottal_flow_plots} shows the glottal flow obtained by inverse filtering, and those obtained by the asymmetric model with the parameters estimated by our ADLES method.
We observe a consistent match, showing the accurateness of our estimations.
Figure~\ref{fig:phase_plots} shows phase portraits of the right and left vocal folds obtained with our ADLES method.
We observe distinctive attractor behaviors for different types of pathologies.
Table~\ref{tab:accuracies} shows the results of deducing voice pathologies by simple thresholding of parameter ranges.
It validates that with our ADLES method, we can accurately estimate model parameters and phase space behaviors, and further use them to classify voice pathologies.
Specifically, the ranges of model parameters in each row of Table~\ref{tab:accuracies} correspond to regions in the bifurcation diagram in Figure~\ref{fig:bifurcation_plots}.
Each region has distinctive attractors and phase entrainment, representing distinct vocal fold behaviors, and thereby indicating different voice pathologies.
By extracting the phase trajectories for the speech signal and thereby the underlying system parameters, the ADLES algorithm is able to place the vocal-fold oscillations in the speech on the bifurcation diagram, and thus deduce the pathology.
% \subsection{Deducing vocal pathologies}
% \textbf{\textit{Deducing vocal pathologies}}
\begin{table}[h!]
\centering
\caption{\footnotesize Parameters obtained and pathologies identified through ADLES}
\label{tab:accuracies}  
\scalebox{0.9}{
\begin{tabular}{c c >{\hspace{0cm}\footnotesize}l >{\hspace{0cm}\footnotesize}l c}
\toprule
$\Delta$ & $\alpha$ & Phase Space Behavior & Pathology & Accuracy \\
\midrule
 $<0.5$ & $>0.25$ & 1 limit cycle, $1:1$ entr. & Norml. & 90\% \\
 $\sim 0.6$ & $\sim 0.35$ & 1 limit cycle, $1:1$ entr. & Neopl. & 82\% \\
 $\sim 0.6$ & $\sim 0.3$ & 2 limit cycles, $1:1$ entr. & Phntr. & 95\% \\
 $\sim 0.85$ & $\sim 0.4$ & toroidal, $n:m$ entr.  & VclPs. & 89\% \\
\bottomrule
\end{tabular}
}
\vspace{-0.5cm}
\end{table}

\section {Conclusions}\label{sec:concl}
The oscillatory dynamics of vocal folds provides a tool to analyze different phonation phenomena, which in turn characterize different types of voice disorders.
In this paper, we have proposed an ADLES method to promote accurate and efficient recovery of the parameters of an asymmetric vocal folds model directly from the speech signal.
This allows us to correctly solve for the oscillatory dynamics of the vocal folds for the specific speech signal.
But more importantly, the parameters estimated for the model directly allow us, through the bifurcation map, to predict voice pathology.
Moreover, the ADLES method significantly alleviates the difficulty of obtaining actual measurements of vocal fold displacements in clinical settings. It can thus be a valuable aid in the diagnosis of different voice pathologies.

%\subsection{Inverse Filtering and Glottal Flow Estimation}
%Vocal tract loading
%(Lip radiation)
%Inverse filter
 
% \newpage
%\appendices
% \bibliographystyle{IEEETran}
\bibliographystyle{IEEEbib}
\bibliography{references}

\end{document}